\documentclass[a4paper,11pt]{article}
\usepackage{pos}

\title{NNPDF4.0 aN$^3$LO PDFs with QED corrections}

\author[a]{Andrea Barontini}
\author[a]{Niccolo Laurenti}
\author*[b,c]{Juan Rojo}

\affiliation[a]{Tif Lab, Dipartimento di Fisica, Universit\`a di Milano and
      INFN, Sezione di Milano, Via Celoria 16, I-20133 Milano, Italy}

\affiliation[b]{Nikhef Theory Group, Science Park 105, 1098 XG Amsterdam, The Netherlands}
\affiliation[c]{Department of Physics and Astronomy, Vrije Universiteit, NL-1081 HV Amsterdam, The Netherlands}

\emailAdd{j.rojo@vu.nl}

\abstract{We review recent progress in the global determination of the collinear parton distributions (PDFs) of the proton within the NNPDF framework.
This progress includes NNPDF4.0 variants with QED effects and with missing higher order uncertainties (MHOUs) via the theory covariance matrix formalism, as well as PDFs based on QCD calculations at approximate N$^3$LO (aN$^3$LO) accuracy. 
We present the combination of these theoretical developments resulting into NNPDF4.0 aN$^3$LO variants accounting for QED corrections and with a photon PDF.
We compare these aN$^3$LO QED PDFs with analogous results obtained by the MSHT group, and briefly quantify their implications at the level of  representative LHC cross-sections.}

\FullConference{31st International Workshop on Deep Inelastic Scattering (DIS2024)\\
 8–12 April 2024\\
Grenoble, France\\}

\usepackage{url}
\usepackage{booktabs,multirow}

\newcommand{\lp}{\left(}
\newcommand{\rp}{\right)}

\begin{document}

\maketitle

\paragraph{Introduction.}
The precise determination of the quark and gluon substructure of the proton, as encoded  by the parton distribution functions (PDFs)~\cite{Gao:2017yyd,Kovarik:2019xvh}, is a central ingredient of the LHC physics program.
This sensitivity to the proton structure is highlighted by recent high-precision measurements of Standard Model (SM) parameters at the LHC~\cite{Amoroso:2022eow}, where the achieved precision is limited by PDF-related uncertainties.
Relevant analyses include the measurements of the strong coupling $\alpha_s(m_Z)$ from the $Z$ $p_T$ spectrum from ATLAS~\cite{ATLAS:2023lhg}, of the effective leptonic mixing angle
$\sin^2\theta_{\rm eff}^\ell$ from CMS~\cite{CMS:2018ktx,CMS:2024aps}, as well as the measurements of the $W$ boson mass $m_W$ from ATLAS~\cite{ATLAS:2024erm} and from the CDF experiment at the TeVatron~\cite{CDF:2022hxs}.
Similar considerations apply to searches of new physics in the high-energy tails of differential distribution, as demonstrated with the Drell-Yan forward-backward asymmetry as a case study~\cite{Ball:2022qtp} as well as by simultaneous PDF and SMEFT analyses of high-$p_T$ data from the LHC~\cite{Greljo:2021kvv,Kassabov:2023hbm}.

In this contribution we summarise recent progress within the NNPDF global analysis framework.
This progress has focused on extending the NNPDF4.0 determination in three complementary directions: QED corrections, missing higher
order uncertainties (MHOUs) by means of the theory covariance matrix approach, and PDFs determined at approximate N$^3$LO (aN$^3$LO) accuracy in the QCD coupling expansion.
Here, for the first time, we present the  combination of these  theoretical developments into
NNPDF4.0 aN$^3$LO sets, with and without MHOUs, accounting for QED corrections and with a photon PDF.
We also quantify the impact of this combination both at the level of PDFs and of benchmark LHC cross-sections, and compare them with their counterparts obtained by the MSHT group. 
These combined PDF sets are made available via the {\sc\small LHAPDF} interface~\cite{Buckley:2014ana} as well as via the NNPDF Collaboration website.\footnote{
\url{https://nnpdf.mi.infn.it/for-users/unpolarized-pdf-sets/}}

\paragraph{NNPDF4.0 with QED corrections.}
The original NNPDF4.0 determination~\cite{NNPDF:2021njg,NNPDF:2021uiq} did not include neither QED corrections to the DGLAP evolution nor a photon PDF $\gamma(x,Q^2)$. 
Building up on the approach deployed for NNPDF3.1QED~\cite{NNPDF:2017mvq,Bertone:2017bme} and based on the LuxQED formalism~\cite{Manohar:2017eqh}, we have recently presented the NNPDF4.0QED analysis~\cite{NNPDF:2024djq}.
This variant of NNPDF4.0 is based on the new theory pipeline~\cite{Barontini:2023vmr} of the NNPDF framework and assembled around the {\sc\small EKO}~\cite{Candido:2022tld}, {\sc\small YADISM}~\cite{Candido:2024rkr}, and {\sc\small PineAPPL}~\cite{Carrazza:2020gss} programs.
The NNPDF4.0QED variants are based on QCD$\otimes$QED DGLAP evolution accounting for $\mathcal{O}\lp \alpha\rp$, $\mathcal{O}\lp \alpha\alpha_s\rp$ and $\mathcal{O}\lp \alpha^2\rp$ corrections implemented and benchmarked in {\sc\small EKO}.
In~\cite{NNPDF:2024djq} we investigate the impact of QED effects on NNPDF4.0, compare our results to other recent PDF sets that include $\gamma(x,Q^2)$, and quantify the impact of photon-initiated processes and electroweak corrections on a variety of LHC processes, finding that they can reach the several-percent level at the high-energy tail of differential distributions. 

\paragraph{NNPDF4.0 with MHOUs.}
While most PDF determinations consider only experimental uncertainties when constructing the covariance matrix used for the $\chi^2$ loss function, the precision of available and future measurements is such that theoretical uncertainties entering the cross-section predictions cannot be neglected anymore.
One of the dominant sources of theory uncertainties are those associated to the missing higher orders (MHOs) in the QCD perturbative expansion.
A first systematic study of MHOUs within a global NNPDF analysis was carried out in~\cite{NNPDF:2019ubu,NNPDF:2019vjt}, based on the theory covariance matrix formalism, where MHOU and their correlations are parameterised in terms of nuisance parameters determined from scale variations.
In this way, MHOUs are accounted for
on a comparable footing as experimental errors. 
This initial study was restricted to NLO, and confirmed that MHOUs are potentially important given the accuracy of modern PDF fits.
Thanks to the new NNPDF theory pipeline mentioned above, we have been able to produce variants of the NNPDF4.0 NNLO global fit which include MHOUs for all input processes~\cite{NNPDF:2024dpb}.
This analysis finds that MHOU enable an improved overall consistency of the PDF determination, with an ensuing moderate reduction of PDF uncertainties at NNLO.

\paragraph{NNPDF4.0 at aN$^3$LO accuracy.}
Calculations of hard-scattering cross-sections at $\mathcal{O}(\alpha_s^3)$ (relative to the Born term) in the strong coupling (N$^3$LO) have become available for a rapidly growing set of hadron collider processes such as inclusive and differential Higgs and Drell-Yan production, see~\cite{Caola:2022ayt}
for an overview, in addition to the deep-inelastic scattering coefficient functions which have been know for a long time. 
To obtain predictions for LHC observables with this same accuracy, partonic cross-sections need to be combined with PDFs determined also at N$^3$LO. 
These N$^3$LO PDFs should be determined by comparing to experimental data theory
predictions computed also at N$^3$LO in the QCD expansion.

With this motivation, in~\cite{NNPDF:2024nan} we presented a first NNPDF fit based on approximate N$^3$LO (aN$^3$LO) calculations.
This result required constructing an approximation to the N$^3$LO splitting functions and anomalous dimensions reproducing all available information from fixed-order computations and from small and large-$x$ resummation; extending the FONLL general-mass scheme for DIS structure functions~\cite{Forte:2010ta} to N$^3$LO; and accounting both for the remaining MHOUs and for the incomplete higher order uncertainties (IHOUs) associated to the derivation of aN$^3$LO expressions for the splitting functions and the massive DIS coefficient functions. 
The NNPDF4.0 aN$^3$LO analysis of~\cite{NNPDF:2024nan} demonstrates that, within the NNPDF framework, aN$^3$LO PDFs are consistent with their NNLO counterparts, that  aN$^3$LO corrections improve the description of the global dataset, and that the perturbative convergence of Higgs and Drell-Yan inclusive cross-sections is enhanced when consistently combining aN$^3$LO PDFs with N$^3$LO partonic matrix elements.
We also find that, as expected, MHOUs on the PDFs become smaller with the increase of perturbative order.

\begin{table}[t]
  \centering
  \small
   \renewcommand{\arraystretch}{1.5}
   \begin{tabular}
   {|c|c|c|c|}
     \toprule
  Fit ID  &  Perturbative accuracy    & Theory cov. mat.      \\
  \midrule
      {\tt NNPDF40\_nnlo\_as\_01180\_qed\_mhou}  &  NNLO$_{\rm QCD}\otimes$NLO$_{\rm QED}$ & MHOU$_{\rm 7pt}$  \\
      \midrule
       {\tt NNPDF40\_an3lo\_as\_01180\_qed }  &  aN$^3$LO$_{\rm QCD}\otimes$NLO$_{\rm QED}$ & IHOU+MHOU$_{\rm 3pt}$  \\
        \midrule
       {\tt NNPDF40\_an3lo\_as\_01180\_qed\_mhou  }  &  aN$^3$LO$_{\rm QCD}\otimes$NLO$_{\rm QED}$ & IHOU+MHOU$_{\rm 7pt}$  \\
 \bottomrule
   \end{tabular}
   \vspace{0.1cm}
   \caption{\small The settings of the three combined PDF sets presented in this contribution.
\label{table:fit_settings} }
\end{table}

\paragraph{Combining aN$^3$LO, MHOUs, and QED corrections in NNPDF4.0.}
Given that the three theoretical developments outlined above  (aN$^3$LO corrections, MHOUs, and QED effects) are separately beneficial for the precision and/or the accuracy of the fitted PDFs, it is only natural to unify them into a combined global PDF determination, along the lines of the recent study by the MSHT group~\cite{McGowan:2022nag, Cridge:2023ryv}.
Therefore, benefiting from the flexibility of the new NNPDF theory pipeline, we present here for the first time new variants of the NNPDF4.0 global analysis combining these three theory developments in various ways.
Specifically, we have produced a variant of NNPDF4.0QED NNLO with MHOUs, and then two variants of NNPDF4.0 aN$^3$LO with QED corrections, either with or without MHOUs, see Table~\ref{table:fit_settings}.
Recall that, as discussed in~\cite{NNPDF:2024nan}, in the aN$^3$LO fits we always include 3-point MHOUs in the hard-scattering cross-sections of hadronic processes to compensate for the missing N$^3$LO $K$-factors.

\begin{figure}[h]
    \centering
    \includegraphics[width=0.86\textwidth]{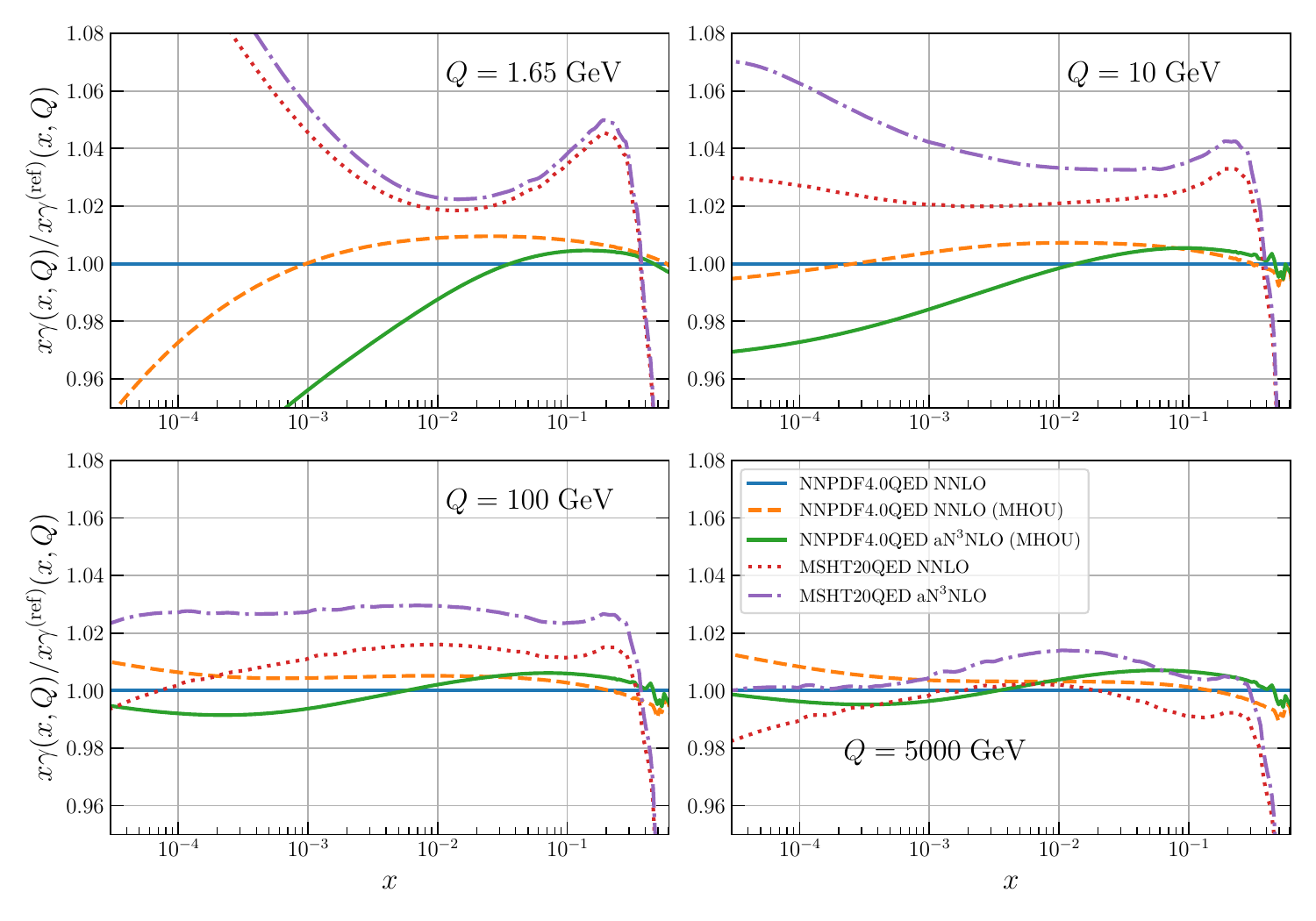}
    \caption{The central values of the photon
    PDF $\gamma(x,Q^2)$ in the NNPDF4.0 and MSHT20 QED sets, both for the
    NNLO and the aN$^3$LO fits.
    For the NNPDF4.0QED fit we also include the variant with MHOUs in the covariance matrix.
    Results are shown normalised to the central value of NNPDF4.0QED NNLO.
    }
    \label{fig:QEDphoton-qdep-ratios}
\end{figure}

We first assess the impact of MHOUs and of aN$^3$LO corrections in the resulting photon PDF $\gamma(x,Q^2)$.
Fig.~\ref{fig:QEDphoton-qdep-ratios} compares the central values of the photon PDF in the NNPDF4.0 and MSHT20 QED sets, both for the NNLO and the aN$^3$LO fits.
For the NNPDF4.0QED fit, we also include the variant with MHOUs in the covariance matrix: it is clear that MHOUs have a very small effect on the photon PDF, unless one goes to very low scales.
Concerning the effect of aN$^3$LO corrections, in NNPDF4.0 they lead to an increase of $\gamma(x,Q^2)$ as compared to the NNLO fit by up to 1\% for $x\ge 0.01$, and the same trend is found for their MSHT20 counterparts.
For a smaller $x$ values, aN$^3$LO effects suppress the NNPDF4.0 photon while instead they enhance the MSHT20 photon. 

\begin{figure}[h]
    \centering
    \includegraphics[width=0.95\textwidth]{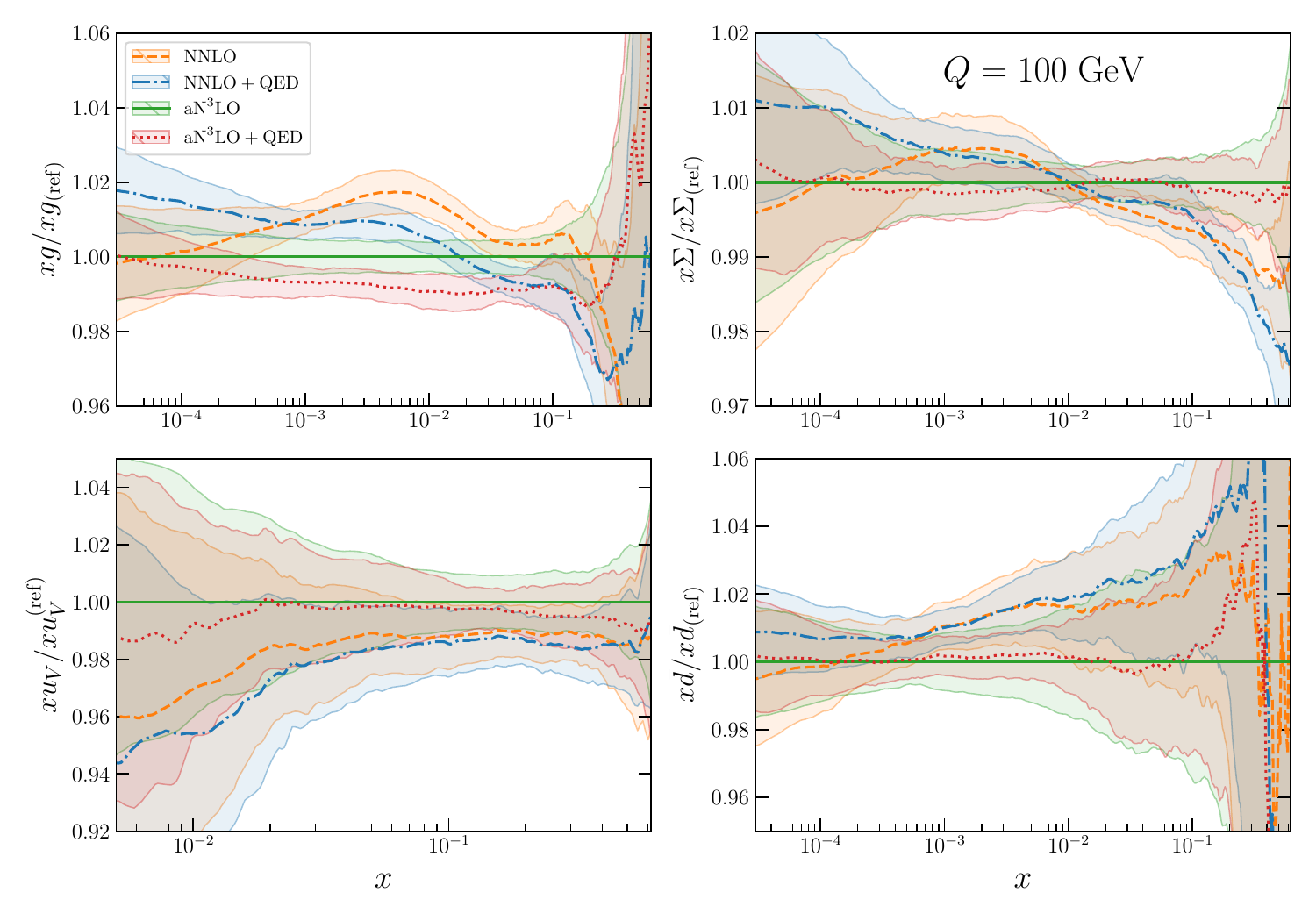}
    \caption{Comparison of the NNPDF4.0 NNLO and aN$^3$LO fits, with and without QED corrections, at $Q=100$ GeV and
    normalised to the central value of
    the aN$^3$LO set.
    Bands indicate the 68\% CL uncertainties.
    }
    \label{fig:PDFs-n3loQED-q100gev}
\end{figure}

We next compare in Fig.~\ref{fig:PDFs-n3loQED-q100gev} the NNPDF4.0 NNLO and aN$^3$LO fits, with and without QED corrections.
We display the gluon, quark singlet, up valence, and antidown PDFs at $Q=100$ GeV.
The qualitative impact of the QED corrections on the quark and gluon PDFs is the same in the NNLO and aN$^3$LO fits.
The largest impact is associated to the gluon, where QED effects lead to an overall decrease of around 1\% as a consequence of the photon PDF contributing to the momentum sum rule.
For the quark PDFs, the impact of QED effects in the aN$^3$LO fit is small, but nevertheless consistent with the results the corresponding NNLO fits.

\begin{figure}[h]
    \centering
    \includegraphics[width=0.313\textwidth]{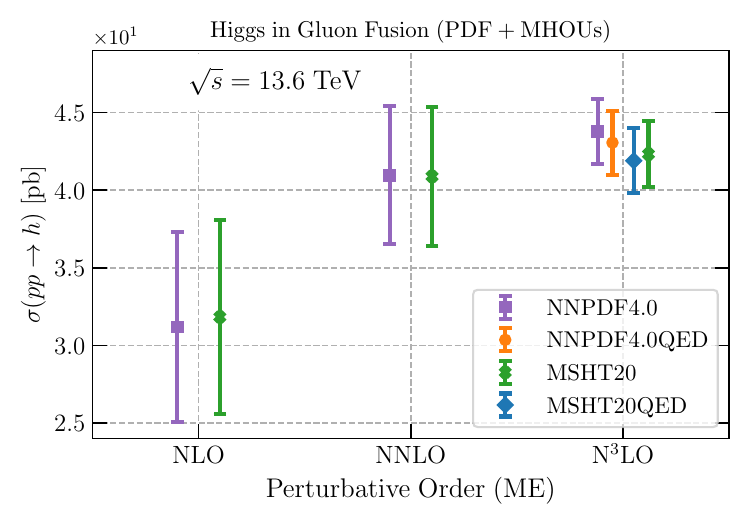}
    \includegraphics[width=0.33\textwidth]{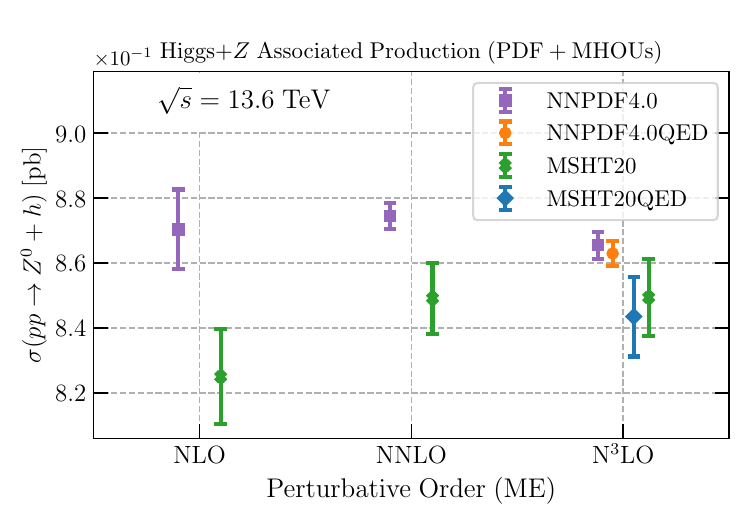}
    \includegraphics[width=0.32\textwidth]{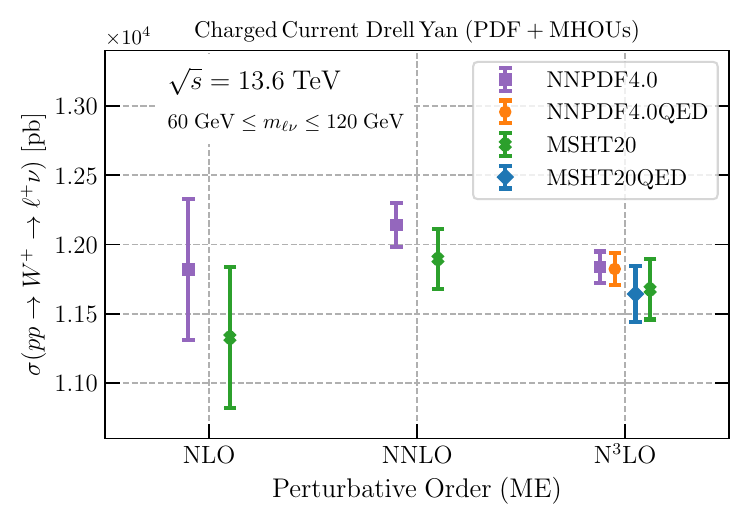}
    \caption{The perturbative convergence 
    of Higgs production in gluon fusion and in associated production with $Z$ bosons, and $W^+$ production at the LHC $\sqrt{s}=13.6$ TeV.
    For the N$^3$LO calculations, we also display predictions based on the QED variants of the NNPDF4.0 and MSHT20 aN$^3$LO sets.
    }
    \label{fig:LHCpheno}
\end{figure}

One can also briefly assess the implications of the combined NNPDF4.0 sets listed in Table~\ref{table:fit_settings} for LHC phenomenology.
To this end, in Fig.~\ref{fig:LHCpheno} we revisit the perturbative convergence analysis of representative inclusive cross-sections considered in~\cite{NNPDF:2024nan} at the LHC with $\sqrt{s}=13.6$ TeV.
We display Higgs production in gluon fusion, in associated production with $Z$ bosons, and $W^+$ production at NLO, NLO, and N$^3$LO.
For the N$^3$LO calculations, we also display predictions based on the QED variants of the NNPDF4.0 and MSHT20 aN$^3$LO sets.
For this comparison, one observes that the qualitative impact of adding QED corrections to the aN$^3$LO sets is the same in the two determinations, namely a moderate reduction in the central values of the cross-sections by an amount well contained within the theory uncertainty band.

\paragraph{Additional NNPDF progress.}
Complementing the developments summarised above, other studies by the NNPDF Collaboration have  either been presented in the past months or will be completed soon.
To begin with, following our study of the total intrinsic charm (IC) content of the proton~\cite{Ball:2022qks}, we investigated the possible asymmetry between intrinsic charm and anticharm quarks.
The crucial observation is that there is no perturbative mechanism in QCD that can generate such an asymmetry between charm and anticharm quarks, and thus its observation represents the ultimate smoking gun for IC in the proton.
A global analysis of hard scattering data indicates a preference for a non-zero, positive intrinsic charm asymmetry~\cite{NNPDF:2023tyk}, although its significance is smaller than for the total component.
To either confirm or falsify this IC asymmetry suggested by current data, in~\cite{NNPDF:2023tyk} we propose two dedicated observables: $Z+$charm production at LHCb at the HL-LHC and $F^c_2$ at the upcoming Electron Ion Collider (EIC)~\cite{AbdulKhalek:2021gbh}.
In both cases, tagging the charge of the final-state charm quark enables the calculation of asymmetries sensitive to the valence IC.

Second, Monte Carlo (MC) event generator programs modelling LHC processes often require PDFs sets with additional constraints as compared to the baseline fits, such as  positive-definite  PDFs down to $Q\sim 1$ GeV, smooth and physical extrapolation down to very small- and large-$x$ as well as small $Q^2$ values, and the inclusion of a photon PDF enabling event generation in the presence of electroweak corrections.
These so-called ``MC PDFs'' also enter tunes of non-perturbative QCD physics, such as the underlying event and multiple parton interactions, necessary for the description of LHC collisions. 
With this motivation we are developing LO, NLO, and NNLO variants of NNPDF4.0 with tailored settings for LHC event generators.
These NNPDF4.0MC sets make possible a reliable description of high-$p_T$ cross-sections as well as, subsequently to a dedicated tuning procedure, of processes sensitive to low-$p_T$ QCD phenomena, while minimising the differences with respect to the baseline PDF sets.

Third, new methodologies are being studied with the aim to stress-test the NNPDF fitting methodology.
These efforts include bypassing neural networks by parametrizing PDFs in terms of Gaussian Processes followed by Bayesian inference~\cite{Candido:2024hjt}, the development of improved hyperparameter optimisation procedures based on the features of the complete PDF probability distribution (and not only of the first moments as in~\cite{NNPDF:2021njg}), as well as the application of closure tests to the determination of PDF-related SM parameters such as the strong coupling constant.

\paragraph{Outlook.}
The progress along several directions in the NNPDF analysis framework outlined in this contribution contributes to the development of NNPDF4.1, an updated global fit to be released in 2025. 
In addition to accounting for QED, MHOUs, and aN$^3$LO effects, the NNPDF4.1 determination will include significantly more data, in particular LHC high-precision measurements based on the complete Run II luminosity.
Thanks to extending the {\sc\small PineAPPL} formalism to NNLO fixed-order calculations, NNPDF4.1 will achieve a global PDF determination without any reliance on the QCD $K$-factor approximation at NNLO. 
This upcoming release will also benefit from  methodological improvements such as robust hyperparameter optimisation methods constructed from estimators based on the information contained on the full PDF probability distributions.
Furthermore, the implementation of NNPDF4.1 in GPU architectures will result in a significant reduction of the overall computational overheads involved, enabling novel PDF studies to be carried out more efficiently.

\bibliographystyle{utphys}
\bibliography{DIS2024-nnpdf}

\end{document}